\documentclass[manuscript]{acmart}
\AtBeginDocument{%
  \providecommand\BibTeX{{%
    \normalfont B\kern-0.5em{\scshape i\kern-0.25em b}\kern-0.8em\TeX}}}

\setcopyright{acmcopyright}
\copyrightyear{2024}
\acmYear{2024}
\acmDOI{XXXXXXX.XXXXXXX}

\acmConference[ACM REP '24]{ACM REP '24: Proceedings of the 2024 ACM Conference on Reproducibility and Replicability}{June 18--20, 2024}{Rennes, France}
\acmBooktitle{ACM REP '24: Proceedings of the 2024 ACM Conference on Reproducibility and Replicability ,
 June 18--20, 2024, Rennes, France} 
\acmPrice{15.00}
\acmISBN{978-1-4503-XXXX-X/18/06}

\usepackage{graphicx}
\usepackage{float}
\usepackage{subcaption}
\usepackage{xcolor}
\usepackage{wrapfig}

\newenvironment{quoteitalicized}
    {\begin{quote}}
    {\end{quote}}
\newcommand{\quotes}[2]{\begin{quoteitalicized}\small \textit{#1} {#2}\end{quoteitalicized}}

\begin{document}

\title[Reproducibility, Replicability, and Transparency in Research:\\ What 430 Professors Think in Universities across the USA and India]{Reproducibility, Replicability, and Transparency in Research:\\ What 430 Professors Think in Universities across the USA and India}

\author{Tatiana Chakravorti}
\email{tfc5416@psu.edu}
\orcid{1234-5678-9012}
\author{Sai Koneru}
\email{sdk96@psu.edu}
\author{Sarah Rajtmajer}
\email{smr48@psu.edu}
\affiliation{%
  \institution{Pennsylvania State University}
  \city{State College}
  \state{Pennsylvania}
  \country{USA}
  \postcode{16803}
}




\renewcommand{\shortauthors}{Trovato and Tobin, et al.}
\newcommand{\sai}[1]{{\color{blue}{#1}}}
\begin{abstract}
In recent years, 

\end{abstract}






\begin{abstract}
In the past decade, open science and science of science communities have initiated innovative efforts to address concerns about the reproducibility and replicability of published scientific research. In some respects, these efforts have been successful, yet there are still many pockets of researchers with little to no familiarity with these concerns, subsequent responses, or best practices for engaging in reproducible, replicable, and reliable scholarship. In this work, we survey 430 professors from Universities across the USA and India to understand perspectives on scientific processes and identify key points for intervention. 
Our findings reveal both national and disciplinary gaps in attention to reproducibility and replicability, aggravated by incentive misalignment and resource constraints. We suggest that solutions addressing scientific integrity should be culturally-centered, where definitions of culture should include both regional and domain-specific elements. 
\end{abstract}
\maketitle

\section{Introduction}
Reproducibility and replicability have gained significant attention in scientific discourse, deeply intertwined with questions about scientific processes, policies and incentives \cite{schooler2014metascience,maxwell2015psychology,loken2017measurement,shrout2018psychology,national2019reproducibility}.\footnote{We adopt definitions from \cite{national2019reproducibility,nosek2021replicability,pineau2021improving}. \emph{Reproducibility} refers to computational repeatability – obtaining consistent computational results using the same data, methods, code, and conditions of analysis; \emph{replicability} means obtaining consistent results on a new dataset using similar methods.} Initially centered around the social and behavioral sciences, these concerns now span almost all empirical scientific disciplines \cite{baker2016reproducibility}, including artificial intelligence and machine learning \cite{willis2020trust, kapoor2023leakage}. The open science and science of science communities have responded with innovative initiatives aimed at shoring up the entire research workflow, from conception and study design, to data collection and analysis, through to publishing and\cite{nosek2015promoting,nosek2016transparency,obels2020analysis}. 
These efforts have already had important individual and institutional impacts, many of which have been well-documented \cite{silverstein2024guide, mazarakis2020gamification}. For example, the Special Interest Group on Computer-Human Interaction (SIGCHI) now recommends providing supplementary materials for ACM publications to enhance replicability \cite{echtler2018open} and some universities have begun to reward researchers whose work aligns with standards of open science and transparency \cite{schonbrodt2018academic}. 

Despite these promising advances, however, conversations around reproducibility and replicability have predominantly reflected the voices of researchers in the global North and West \cite{mede2021replication, cova2021estimating, vilhuber2020reproducibility, fidler2018reproducibility}. This is concerning for a number of reasons, most primarily because issues of scientific integrity and scientific process are deeply social and contextual. Our work takes an initial step toward inclusion of cultural perspectives through comparative study of researchers in the USA and India. India currently ranks third in research output worldwide, following China and the USA \cite{india2023publications}. 

We conducted a survey-based study involving research faculty from universities in the USA and India. Our aim was to gather the perspectives of scientists across different research disciplines. The survey asked participants about their familiarity with the reproducibility crisis, their confidence in work published within their fields, and the factors they believe contribute to this high or low-confidence research. Additionally, we asked participants to share the institutional and practical challenges they face during their research. We reached out to over $8000$ research faculty members and received a total of $430$ responses. 
The following research questions drive this work:

\begin{itemize}
    \item\textbf{RQ1}: What are researchers' \textbf{awareness and concerns} around reproducibility, replicability, and open science in the USA and India? What are the experiences of researchers in these countries with respect to reproducing or replicating others' studies?
    \item\textbf{RQ2}: How do \textbf{institutional challenges and opportunities} differ between these two countries with respect to reproducibility and replicability? What factors contribute to lack of reproducibility?
    \item\textbf{RQ3}: What \textbf{signals of credibility} do our participants look for when reading and evaluating published work? 
\end{itemize}

\noindent Our findings contribute to the global conversation on scientific integrity, underscoring the need to understand challenges and solutions in cultural context.

\section{Related Work} 
\subsection{Reproducibility and the Replication Crisis}
In recent years, several scientific disciplines, including psychology, medicine, cancer biology, economics, and machine learning, have experienced challenges with reproduction and replication \cite{semmelrock2023reproducibility}. 
This was brought into public eye by a Nature survey reporting that 70\% of its researchers have attempted and failed to reproduce another scientists' experiments, and more than half have been unable to replicate their own \cite{baker20161}. 
Large-scale replication projects in psychology \cite{nosek349corresponding}, economics \cite{camerer2016evaluating}, sociology \cite{camerer2018evaluating}, biology \cite{errington2014open} and beyond have turned up disappointing results.  Various factors have been suggested as contributing to this replication crisis, including selective reporting \cite{rowe2023recommendations}, insufficiently detailed methodology descriptions, p-hacking \cite{head2015extent}, poor theoretical design, small sample size, coding mistakes, and the unavailability of code and data \cite{miyakawa2020no}. Major challenges to reproducible and replicable research practices include misaligned incentives \cite{nosek2012scientific, raff2023does} and failures of the peer review process \cite{guttinger2019characterizing}. These issues can be de-motivating for researchers and lead them to prioritize publication quantity over quality. In response, researchers have promoted the adoption of open science practices, e.g., sharing of research artifacts \cite{mcgrath2018data} and the integration of reproducible and replicable practices into academic teaching and education \cite{fund2023we}.

\subsection{Open Science}
Concerns about reproducibility and replicability are closely related to principles and practices of open science \cite{miguel2014promoting, bradley2020reducing, aguinis2019transparency, raff2023siren}. The UNESCO Recommendation on Open Science defines open science, sweepingly, as ``an inclusive construct that combines various movements and practices aiming to make multilingual scientific knowledge openly available, accessible and reusable for everyone, to increase scientific collaborations and sharing of information for the benefits of science and society, and to open the processes of scientific knowledge creation, evaluation, and communication to societal actors beyond the traditional scientific community. It comprises all scientific disciplines and aspects of scholarly practices, including basic and applied sciences, natural and social sciences, and the humanities, and it builds on the following key pillars: open scientific knowledge, open science infrastructures, science communication, open engagement of societal actors and open dialogue with other knowledge systems'' \cite{unesco2021unesco}. In this context, researchers have begun to scaffold clear and specific practices that align with open science; chief amongst them is the notion of transparency. Transparent research practices include sharing data and code, comprehensive detailing of methodologies, and clear identification of theoretical foundations \cite{nosek2015promoting,knottnerus2016promoting}. Researchers have found that making code available has a positive correlation with increased citations \cite{raff2023does}. The specific character of best practices, of course, varies across disciplines \cite{talkad2020transparency}. 
Many fields are working to establish their own norms inspired by open science ideals \cite{errington2021challenges, delios2022examining}. 


\section{Methods}
We take an exploratory, survey-based approach for a 
comparative analysis of researchers' perspectives on reproducibility, replicability, transparency, and open science in India and the USA. 
 
\subsection{Survey recruitment}
We selected 25 universities randomly from the top 100 universities in India based on the National Institutional Ranking Framework (NIRF) \cite{ranking2018national}. Likewise, we selected 25 universities randomly from the top 100 in the USA based on 2023 US News and World Report rankings \cite{us_ranking}. \emph{The full list of universities from which participants were recruited is available on the paper's Github repository} (link below). 

We directly emailed research faculty listed on departmental web pages using email addresses collected via web scraping of Universities' directories. We targeted the following disciplines in the social sciences: economics, political science, education, psychology, sociology, and marketing. From engineering, we targeted: computer science engineering; electrical engineering; electronics engineering (India); and mechanical engineering. In total, we emailed 3942 faculty members in India (910 social sciences, 3032 engineering) and 4400 in the USA (2100 social sciences, 2300 engineering). Our email contained a link to the survey, deployed as a Google Form. A one-time follow-up email was sent to all recipients approximately two weeks later. Participation was voluntary. A total of 430 respondents completed the survey, 169 from India and 261 from the USA. This represents a 4.28\% response rate from India and 5.93\% from USA. 

\subsection{Data collection and analysis}
Our survey protocol asked participants to share their perceptions of the state of reproducibility and open science in their respective academic communities and disciplines, factors they believe contribute to lack of reproducibility and replicability of findings, challenges, and opportunities to promote reproducible research practices. The survey takes approximately 15mins time to complete. \emph{Our complete survey protocol along with the complete set of anonymized survey responses is available on the paper's Github repository: \url{https://github.com/Tatianachakravorti/ACMREP24}.} 

Survey responses were analyzed using descriptive statistics and exploratory data visualizations. Open-ended questions (free text responses) were analyzed using thematic analysis \cite{blandford2016qualitative, terry2017thematic}. This qualitative data analysis approach uses thorough examination of free text responses to uncover patterns and derive themes related to the research questions. Two researchers examined all open-ended responses to identify themes relevant to research questions and establish initial codes. After that, codes were organized into categories based on similarities and relationships. Lastly, we refined these categories, assigned names to each theme, and crafted a conceptual framework to address our research questions.


\section{Findings}



\subsection{Awareness and concern about reproducibility and replicability (RQ1)}
\begin{figure}
     \centering
     \begin{subfigure}[t]{0.32\textwidth}
         \centering
         \includegraphics[width=\textwidth, height=5cm, keepaspectratio]{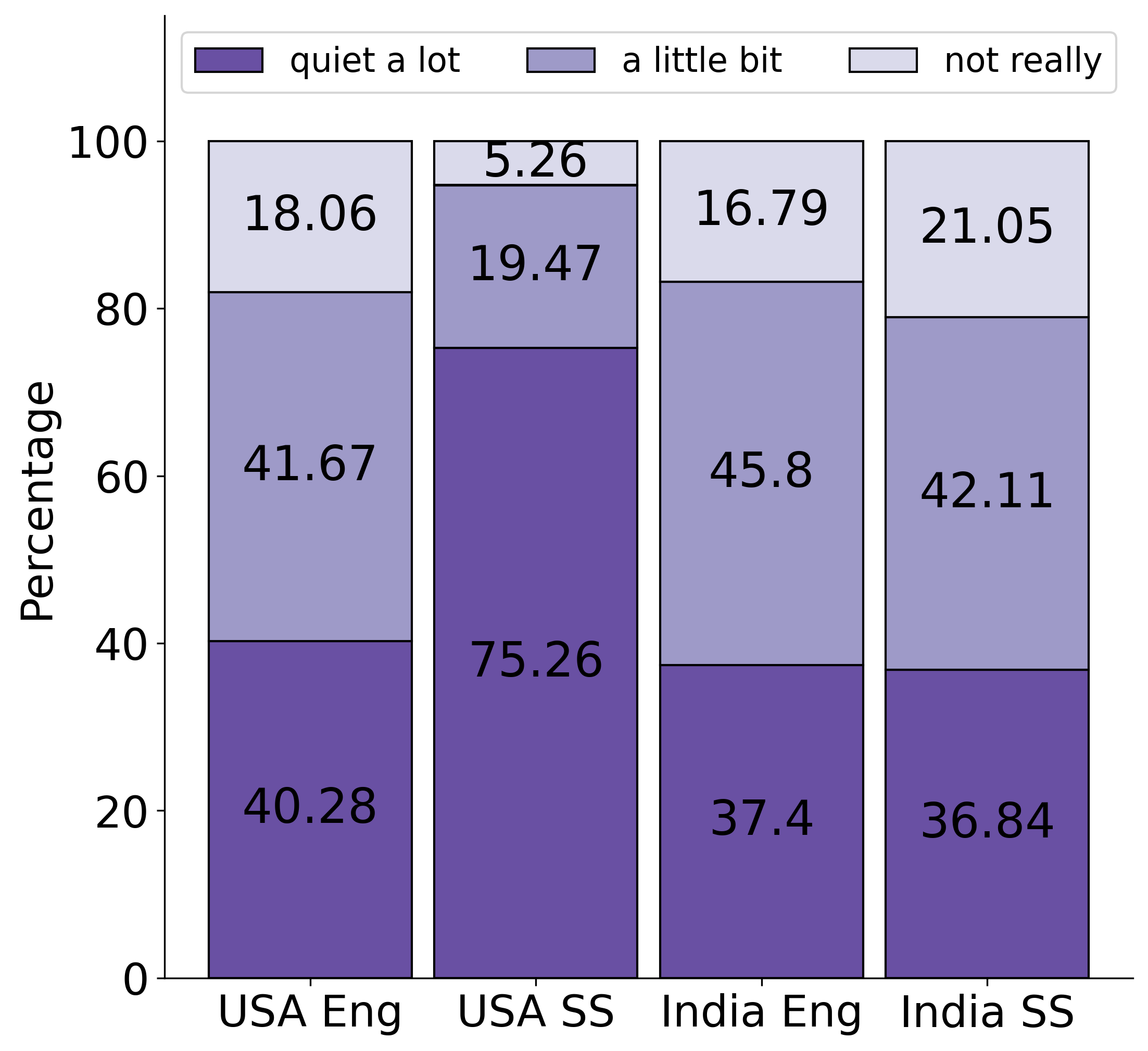}
         \caption{Have you heard much about the "reproducibility crisis" in science?}
         \label{fig:rep_crisis}
     \end{subfigure}
     \hfill
     \begin{subfigure}[t]{0.32\textwidth}
         \centering
         \includegraphics[width=\textwidth, height=5cm, keepaspectratio]{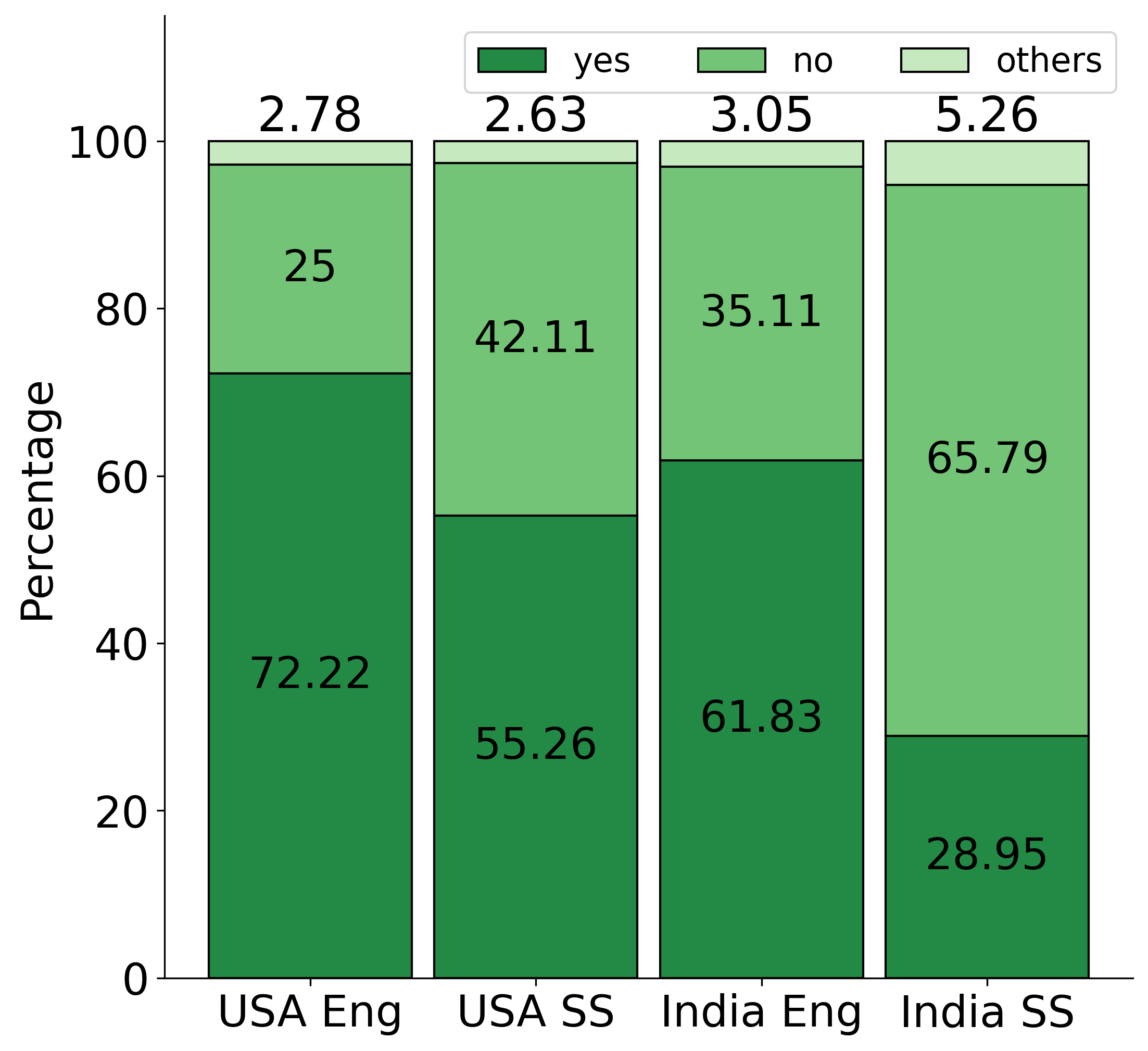}
         \caption{Have you ever tried to repeat a research study someone else published?}
         \label{fig:repeat_others}
     \end{subfigure}
     \hfill
     \begin{subfigure}[t]{0.32\textwidth}
         \centering
         \includegraphics[width=\textwidth, height=5cm, keepaspectratio]{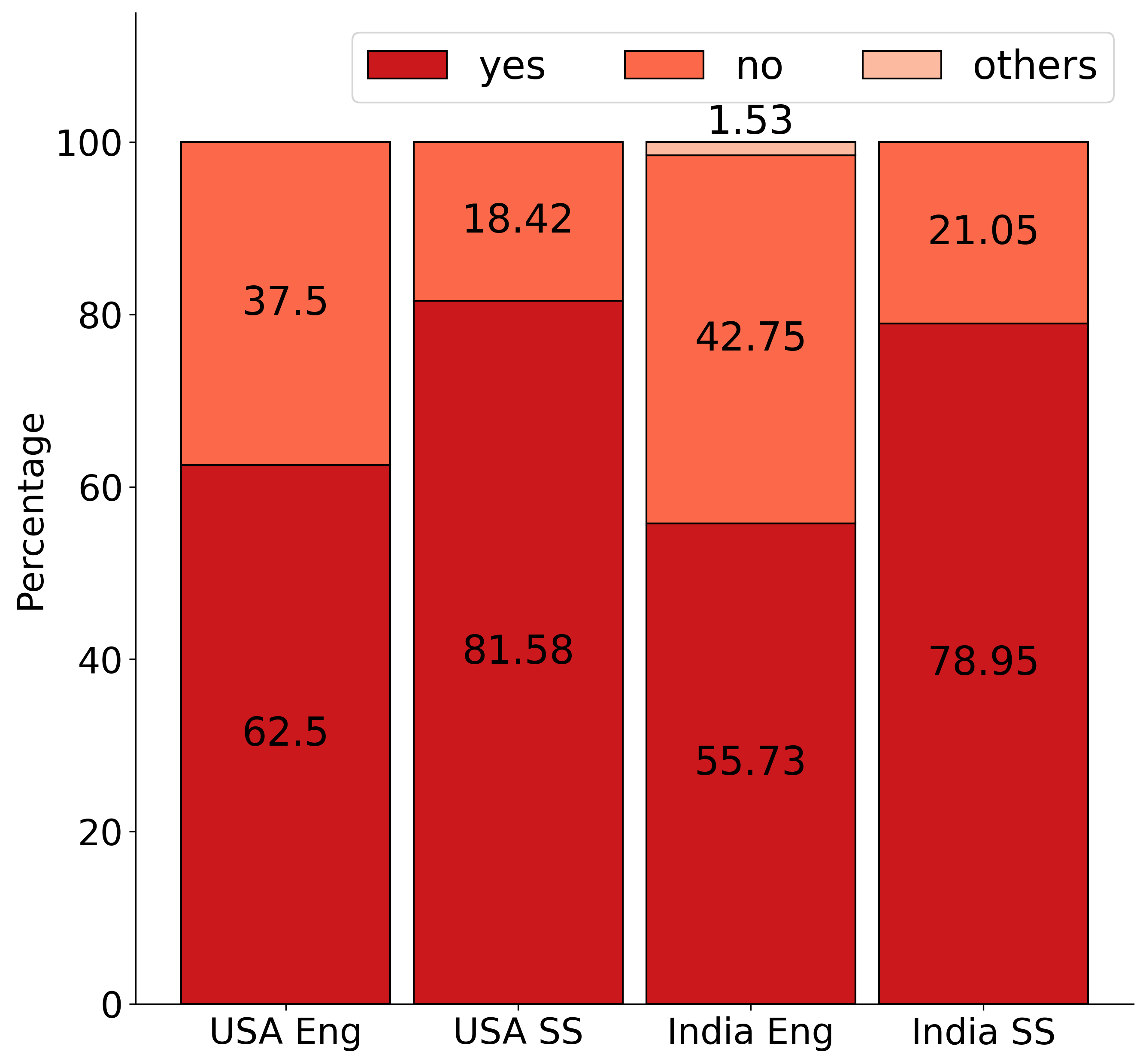}
         \caption{Have you heard much about the "open science" movement?}
         \label{fig:open_science}
     \end{subfigure}
     \\
     \begin{subfigure}[t]{0.32\textwidth}
         \centering
         \includegraphics[width=\textwidth, height=5cm, keepaspectratio]{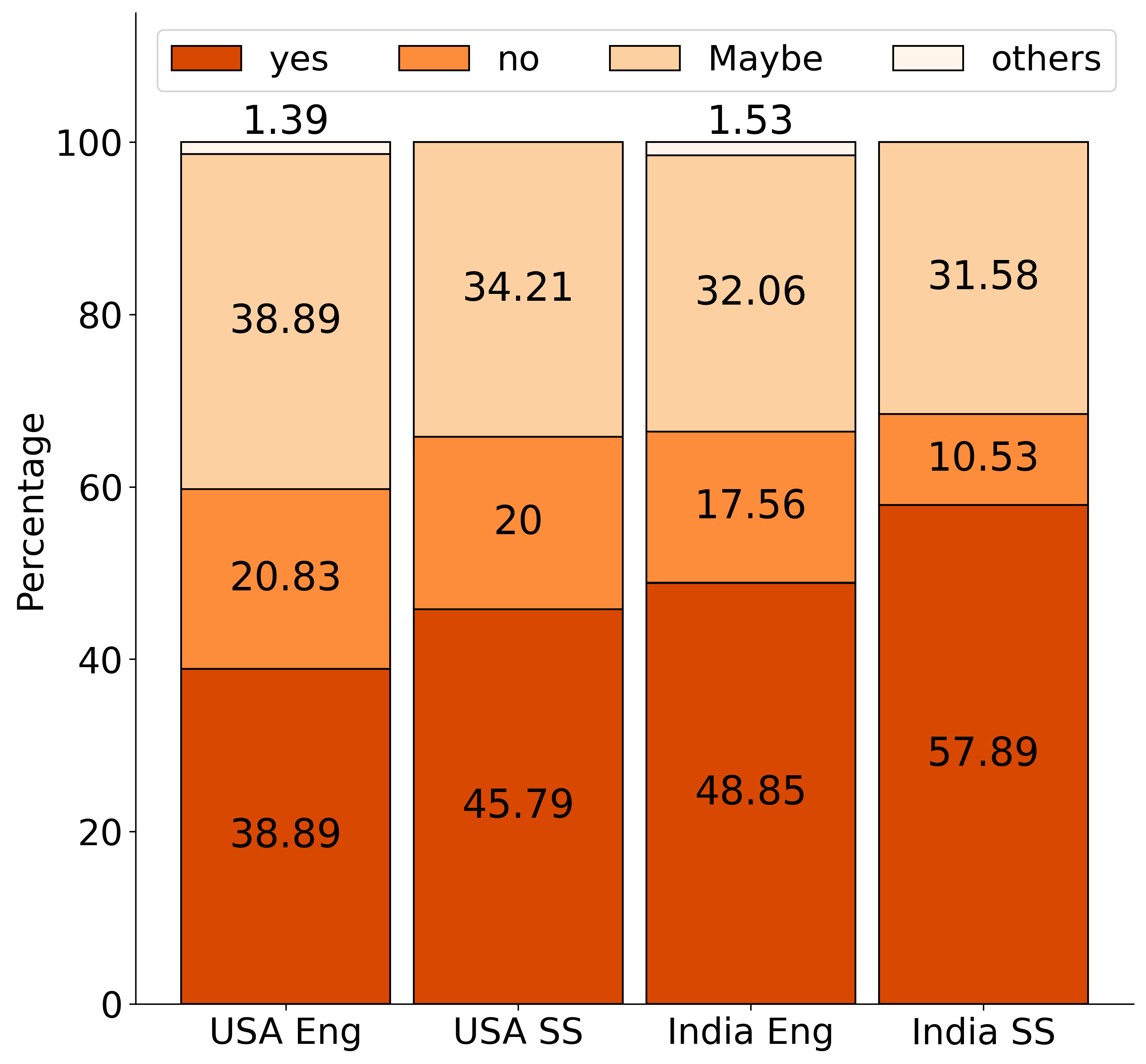}
         \caption{Should reproducibility and open science be a part of undergraduate curriculum?}
         \label{fig:course_work}
     \end{subfigure}
    \hfill
    \begin{subfigure}[t]{0.32\textwidth}
         \centering
         \includegraphics[width=\textwidth, height=5cm, keepaspectratio]{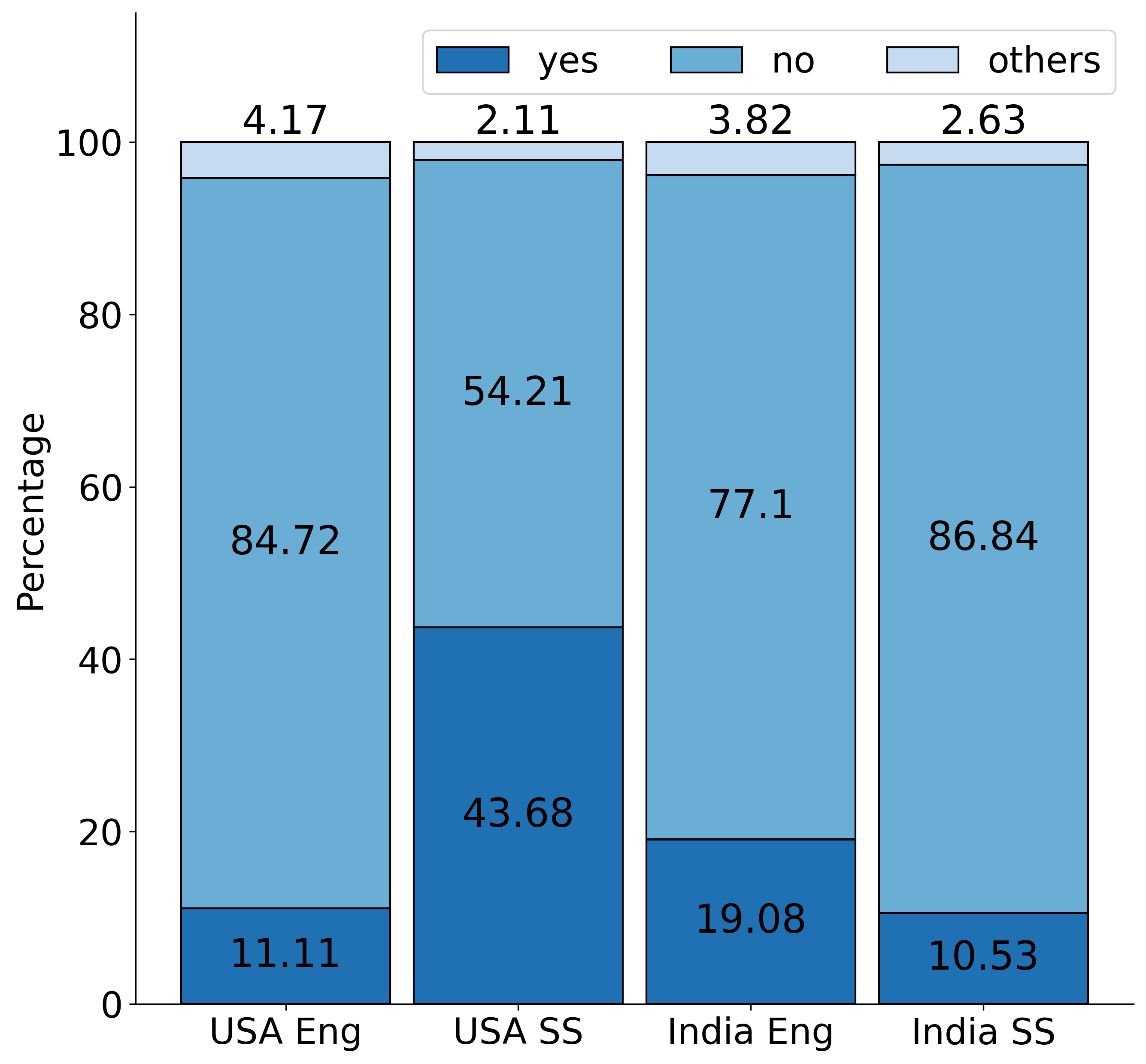}
         \caption{Have you ever pre-registered a study?\\  }
         \label{fig:prereg}
     \end{subfigure}
     \hfill
         \begin{subfigure}[t]{0.32\textwidth}
         \centering
         \includegraphics[width=\textwidth, height=5cm, keepaspectratio]{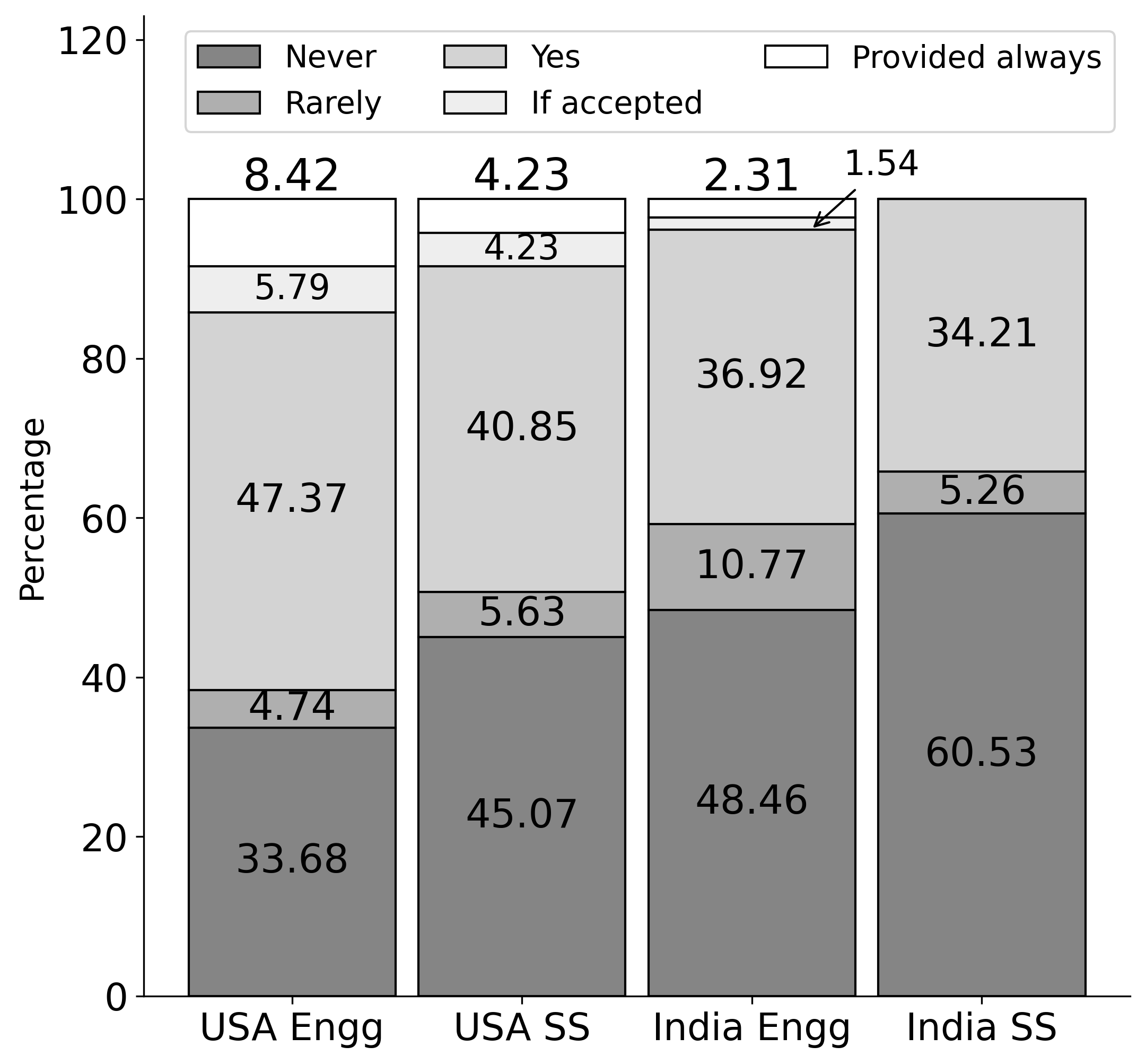}
         \caption{Have you been asked to share research materials during the review process?}
         \label{fig:peerreview}
     \end{subfigure}

        \caption{Survey responses by country and domain (Eng=engineering; SS=social science)}
        \label{fig:survey_results}
\end{figure}

\textbf{Approximately 82\% of surveyed researchers in India indicated some level of familiarity with the reproducibility crisis in science.  In the USA, awareness was over 91\%.} 
Breaking these totals down further, we observe appreciable differences between disciplines. 
More than 94\% of social science researchers in the USA are aware of the reproducibility crisis vs. 81.95\% in engineering. While, in India, this gap is smaller, with 83.2\% of researchers in engineering and 78.95\% in the social sciences endorsing awareness of these concerns (see Figure \ref{fig:rep_crisis}). Additionally, we explored the respondents' perceptions of their peers' awareness of the crisis. \textbf{We find that 27.78\% of participants from India and 17.62\% from the USA believe their peers to be completely \emph{un}aware of the replication crisis.} These statistics underscore 
differences in open discussions about scientific credibility and practice in India and the USA.

\subsubsection{\textbf{Factors contributing to lack of reproducibility}}
Low statistical power was identified as an important concern in social science research.  In fact, 42.63\% of social science participants from the USA and 23.68\% from India identified low statistical power as a contributor to lack of reproducibility. Unavailability of data and code was the primary reason identified amongst engineers from both countries (more than 54\%). Specifically, 59.17\% of Indian researchers identified the unavailability of raw data as a significant obstacle and 51.48\% noted the unavailability of code. Of researchers from the USA, 55.56\% mentioned data unavailability and 40.61\% pointed to code unavailability as a key issue. 

\begin{figure}
\vspace{0.5cm}
  \includegraphics[width=\textwidth]{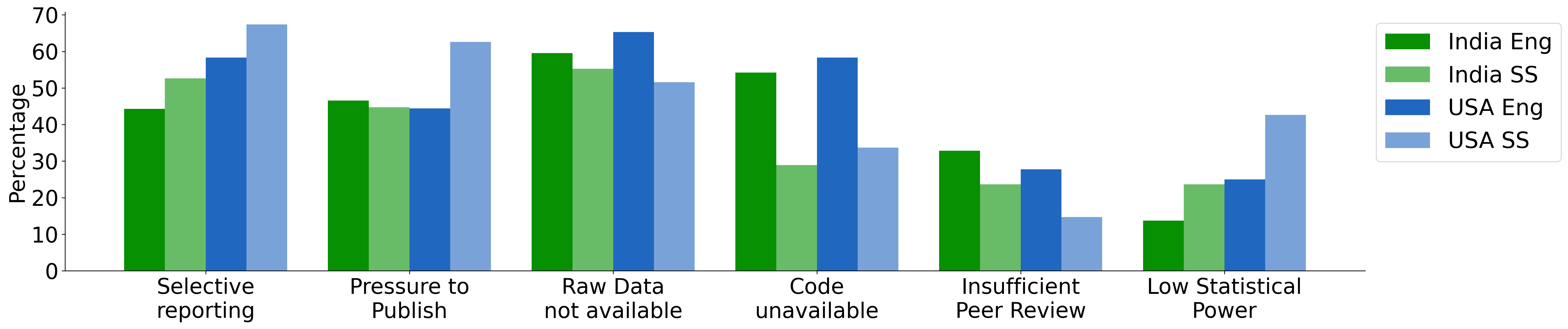}
   \caption{Factors contributing to lack of reproducibility}
   \label{fig:lack_of_rep}
   \vspace{-0.2cm}
\end{figure}

Selective reporting was also a concern, with 64.75\% of American researchers and 46.15\% of Indian researchers recognizing it as a significant issue. Publication pressure was another important factor, with 57.85\% of respondents from the USA and 46.15\% from India citing it. While, insufficient peer review was mentioned by 30.78\% of Indian researchers compared to 18.01\% from the USA (see Figure \ref{fig:lack_of_rep}). 

\subsection{Experiences reproducing and replicating others' findings (RQ1)}

Reproducing and replicating others' work is fundamental to scientific research. Approaches to do so vary based on the nature of research and norms within disciplines. When asked about their experiences replicating others' work, many researchers indicated that they repeat others' experiments before extending them in their own studies. 
\textbf{In India, 54.44\% of respondents indicated that they have tried to replicate others' research, compared to 59.77\% in the USA.} 
Subsequently, researchers who reported having engaged in replication attempts were asked to share insights about their experiences as an open-ended question. \textbf{Only 14.58\% of Indian researchers who reported having tried to replicate others' research reported having obtained affirmative results} with the remaining majority reporting unsuccessful or only partially successful results. \textbf{In comparison, 33.96\% of respondents in the USA who engaged in replication of others' work reported affirmative results.} 
Looking at disciplinary impacts, we find that 55.25\% of social science researchers in the USA have attempted to replicate others' research vs. 28.95\% in India (see Figure \ref{fig:repeat_others}.) In engineering disciplines, 72.22\% of participants from the USA and 61.83\% from India have tried to replicate others' findings. Open-ended responses indicated that this difference may be attributed to differences in access to resources required for replication, e.g., funding, computing. In fact, researchers in both countries reported similar challenges during replication attempts. The most frequently amongst these challenges were resource constraints, specifically, time and money. 
An open-ended question on our survey asked respondents to share their experiences when engaging in reproducing and replicating others' findings. We analysed these responses using thematic analysis, as described above. Following, we detail extracted themes.

\subsubsection{\textbf{Insufficient detail provided in the paper}}
Of respondents who reported that they could only partially replicate existing studies, most mentioned lack of adequate information provided in the paper as the primary reason. 
Participants emphasized the importance of effective documentation for enabling reproducibility, noting that the specific requirements of this documentation varies by domain. For example, one respondent noted the importance of documenting model hyperparameters.
\quotes{I work in the area of applications of deep learning (DL) to IoT. Most of the times, I have observed that the performance results for the DL models are not really replicable. The reason could be the authors don't share all the hyperparameters for model training. However, even after trying with a range of hyperparameter settings, we couldn't replicate the results. This even happens for the A* conference papers. Hope with your findings and the corresponding publications, the researchers will start thinking about sharing the required information to reproduce the results.}{-respondent from India, engineering}

In some cases, a successful reproduction or replication was achieved by contacting the study's authors to fill in missing information, highlighting the importance of cooperation and collaboration. 
\quotes{It went well, we had to contact the authors to get some details that were not available in the paper, but they were responsive. Our results were affirmative.}{-respondent from USA, computer science}
However, other respondents reported instances where authors did not respond to their inquiries. \textbf{Particularly for researchers in India, contacting a study's original authors and receiving a response appears to be more challenging.} While India is increasingly engaged in international networks, the country still faces barriers in this regard, influenced by geopolitical and economic factors. These barriers may play a role in lack of scientific dialogue.
\quotes{Many times the code and data can be obtained from the researchers. However, sometimes they do not respond and do not make the data/code available either. That is quite frustrating really. However, sometimes the easy way out is just to implement ourselves, compare, and then report in the paper, mentioning the code/ data was unavailable.}{-respondent from India, information science}


\subsubsection{\textbf{Conflicting results}}
Even when all relevant information and artifacts were available, some researchers found themselves unable to successfully reproduce published findings, often obtaining conflicting results with those reported in this original work. This experience was pervasive amongst researchers surveyed in both the countries. 
\quotes{We had enough information to repeat the study, a computer design study which was published at a well-known conference, but we got conflicting results.}{-respondent from USA, computer science}

\quotes{Tried to simulate based on the details provided by the author in the paper, but failed many times to reproduce the results shown.}{-respondent from India, electrical Engineering}

\subsubsection{\textbf{Affirmative results}}
Optimistically, many of our participants reported successfully reproducing and replicating findings in the literature, and many of them went on to extend those findings in their own studies. This was particularly the case for our respondents from the USA; who reported having affirmative results when they replicated others' work which is much higher compared the India. Very few researchers mentioned about the affirmative results from India. But overall the statistics highlight that it was not very easy for the Indian researchers. All these percentages are from the group that tried to replicate others' work.
\quotes{Successful. The reproduction was affirmative. I have also had others repeat studies my group has conducted.}{-respondent from USA, computer science}

\subsubsection{\textbf{Hopeful progress}}
Respondents indicated that in the past, reproducing others' work was significantly more challenging due to inadequate documentation and the absence of code and data sharing. However, the situation has considerably improved in step with the open science movement.


\quotes{In three tries the answers vary. Case1 (in 1986) was a very difficult and time-consuming process that reflected the low standards of research process control at the time. Case2 (in 2006) went much better and the outcome was much closer correspondence between the published results and my student's replication of them. Case3 (in 2019) was an exact replication using the code the authors had posted on OSF.}{-respondent from USA, sociology}


\begin{figure}
\vspace{0.5cm}
  \includegraphics[width=\textwidth]{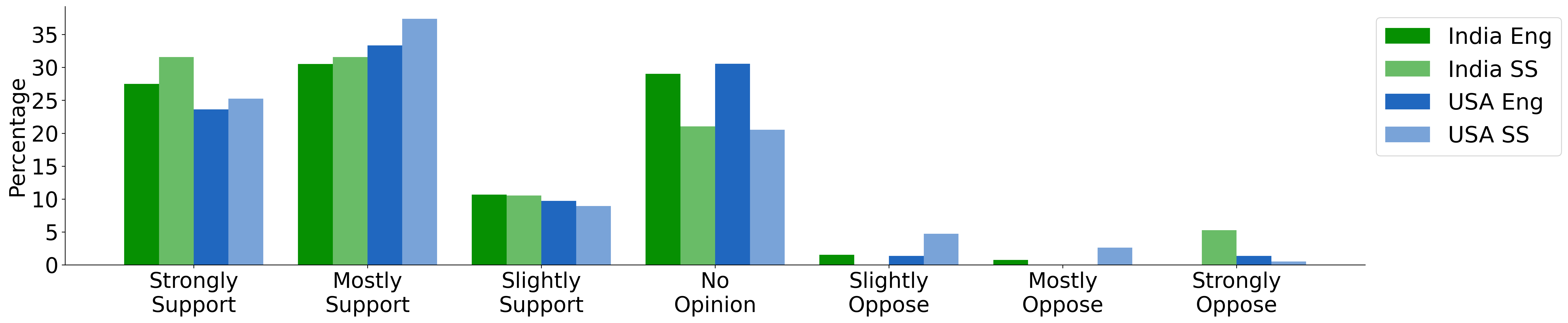}
   \caption{Opinions about the open science movement}
   \label{fig:opinion}
\vspace{-0.2cm}
\end{figure}

\subsection{Attitudes towards open science (RQ1)}
More than half of respondents in both countries reported some awareness of the open science movement and its aims. Specifically, in the USA, 76.25\% of respondents indicated awareness vs. 60.95\% in India. In both countries, social science researchers were more markedly more familiar with the open science discourse than engineers (see Figure \ref{fig:open_science}). 
Overall support for open science principles was assessed on a 7-point Likert scale, ranging from \emph{no support} to \emph{strong support} (see Figure \ref{fig:opinion}). 

\subsubsection{\textbf{Sharing of research artifacts}}
Respondents who indicated \emph{no support} for the open science movement cited concerns about the time and cost associated with sharing data and code. Several respondents compared the sharing of resources to revealing proprietary work before completing a long-term project. They shared they can write multiple papers based on a single dataset or piece of code, and that they were reluctant to share with others to avoid heightened competition.
\quotes{Some code takes years to produce and researchers are still publishing papers from it after one paper is published. Sharing it means that you are essentially giving away your proprietary work before a long-term project is complete.}{-respondent from USA, sociology}

\noindent Additional factors contributing to lack of support for open science include the belief expressed by some researchers that the open access movement lacks inclusiveness and disadvantages less-resourced individuals.  Participants expressed concerns related to the low quality of peer reviews at many open access journals. In addition, researchers engaged in human subjects research expressed difficulty making all aspects of their work openly accessible due to privacy concerns.

\quotes{Human subjects research requires sensitivity for confidentiality and privacy. In public health research there can be powerful opponents who may use data inappropriately. People untrained in statistics and epi could obfuscate important health issues - erodes respect for experts.}{-respondent from USA, social science}

\quotes{If I were to not whole heartedly support it, it might be because open science movement is restricted to elite institutions, led by researchers from  developed countries where the resources available and the challenges faced by the researchers are very different from the ones faced by researchers from the developing countries (participant unavailability, low incentives for participation in research, power failures, lack of lab space, non compliance of participants to protocol despite consent unique to developing countries, some institutes do not have ethics board to approve a study etc).}{-respondent from India, psychology}

We asked our participants whether they had ever been asked to share data files, code, or other materials during peer review of their own work. This question was open-ended and we used thematic analysis to analyze their responses. To explore the role of peer review in promoting reproducibility and replication, researchers were surveyed about their experiences with the peer review process. Specifically, they were asked if they had been requested to share data files, analytic code, or other research-related materials during review. This inquiry was posed as an open-ended question, allowing participants to freely share their thoughts and experiences. Through thematic analysis of their responses, we identified six major themes: Never, Rarely, Yes, provided always, and if accepted (see \ref{fig:peerreview}). 

\quotes{Many top computer vision or machine learning conferences now encourage code submission during paper submission. However, as a reviewer (due to time pressure) I am not willing to check the code by running it. So, I think, ultimately it comes down to reducing the load on the reviewers which, in turn, means training more reviewers to do the job right.}{-respondent from India, engineering}

\begin{wrapfigure}{r}{0.4\textwidth}
 \vspace{-0.5cm}
   \includegraphics[width=0.4\textwidth]{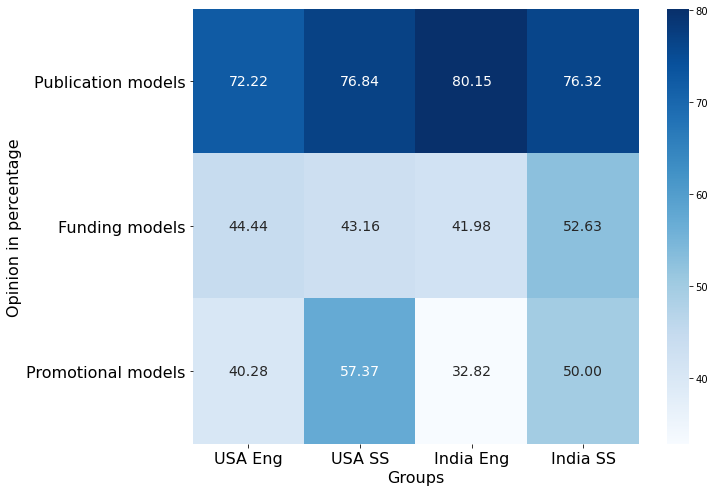}
    \caption{Participants were asked whether changes to publication, funding, or promotional models (inclusive) were needed to advance reproducibility.}
 \vspace{-.5cm}
    \label{fig:changestomodels}
\end{wrapfigure}

\subsubsection{Preregistration}
We explored participants' perspectives on a core open science practice, namely, preregistration. Experience with preregistration varies between India and the USA, with only 17.16\% of researchers in India reporting preregistration compared to 34.48\% in the USA. Notably, in India, only 10.81\% of social science researchers reported having ever preregistered a study, as opposed to 44.74\% in the USA. Notably, however, the percentage of engineering researchers in the USA who have preregistered work was lower at 11.11\% than in India at 21.22\% (see figure \ref{fig:prereg}). The results underscores that ways in which disciplinary culture can be as meaningful, if not moreso, than country-specific norms. 
\quotes{My work does not involve statistical work of the kind implied in this question. We are physical scientists so we don't need pre-registration.}{-respondent from India, mechanical engineering}
A respondent pointed out that preregistration may not be relevant for most studies, further emphasizing varied perceptions among researchers about the utility of preregistration.
\quotes{Preregistration is stupid for the vast majority of studies. If you are doing a high-risk/cost intervention hypothesis test, fine. But very few people are doing that.}{-respondent from USA, sociology}

\subsection{Institutional challenges and opportunities (RQ2)}

As noted, our survey explored obstacles to successful reproduction and replication. These include the unavailability of code or data, unclear explanations of experimental settings, and lack of detailed methodological descriptions. Our survey protocol also allowed respondents to enter "other" factors they perceived to contribute to lack of reproducibility in their field. The majority of these "other" issues were institutional or systemic, related to research culture, norms, incentives, peer review processes, and training.

\subsubsection{Misaligned incentives}
The most commonly noted of these factors across both countries and both fields was \textbf{lack of incentives} for reproducing or replicating others work. Few journals regularly publish reproduction, but rather highlight novelty. This has substantial impact given respondents reporting feeling significant pressure to publish. 
In fact, participants were asked to choose from a range of institutional changes that could help to support reproducible research practices, with the option to select multiple choices. Across both countries, the most commonly selected response was "changes to publication models" (79.3\% of researchers from India and 75.48\% from the USA).
Additionally, 57.37\% social scientists from the USA suggested "changes to promotional models" are required to incentivize best practices. In both countries, engineering researchers reported changes to funding models to be equally or more important than changes to promotional models (see figure \ref{fig:changestomodels}). In this section, we provided the "other" option to make it more flexible and open to the respondents and many of them responded using this option which contributes to one major incentive would be the rigorous peer review process. According to many the peer review is not sufficient to understand replicability and data sharing. Journals and conferences should not accept papers if the code and data are not provided by the authors.
\quotes{Review and publication process should really be focusing on the rigor of the methods, not the significance of the results; with valid and generalizable methods, insignificant/unexpected results are still important, which means we thought it wrong.}{-respondent from USA, psychology}

\subsubsection{Coursework}

A hope of the open science movement is that a global shift toward more rigorous research practices can be achieved through educational efforts targeting the next generation of researchers. We asked participants about their willingness to incorporate reproducibility and open science topics into their coursework. In India, 57.89\% of social science researchers and 48.85\% from engineering believed it is necessary to add open science and replication topics to current course work. In the USA, these numbers were slightly lower; 45.79\% of respondents from social science and 38.89\% from engineering believed this is needed (see figure \ref{fig:course_work}). 

\begin{figure}
    \vspace{.5cm}
  \includegraphics[width=\textwidth]{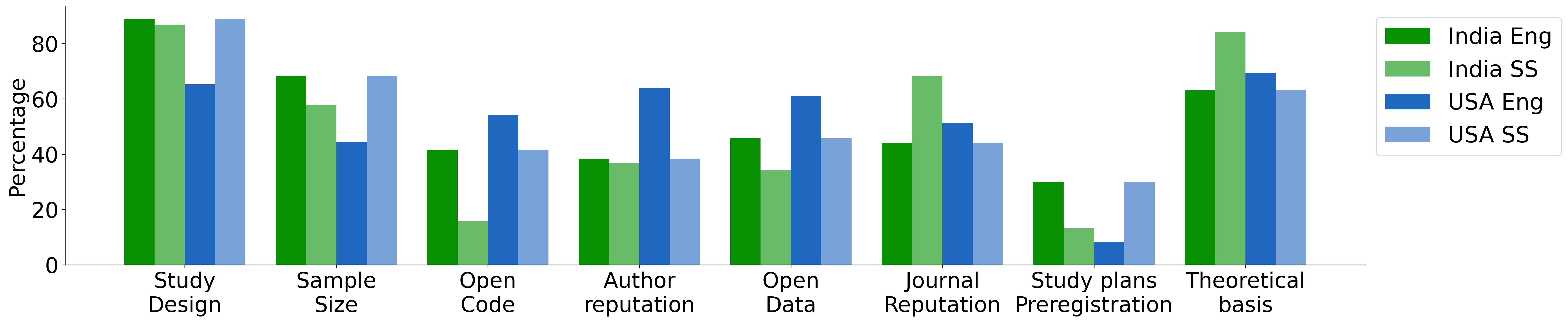}
   \caption{Signals of credibility}
   \label{fig:signals}
    \vspace{-.5cm}
\end{figure}

\subsection{Signals of credibility of published findings (RQ3)}
We sought to understand how our participants evaluate the credibility of published findings when they see them in the literature, e.g., based on journal reputation, whether authors have shared materials, robustness of study design, and similar. We offered a lengthy list of potential signals and also left space for respondents to include their own, understanding that important signals likely vary across domains. 
Study design and theoretical basis were revealed to be the most frequently identified across both countries as signals of credibility of published findings. Engineering researchers tend to look for open sharing of data and code, whereas social science researchers evaluate the sample size of the study population. Author reputation and journal reputation were also identified as significant factors, particularly identified as such by respondents from the USA (see figure \ref{fig:signals}).

\quotes{I typically pay extremely close attention to the detail with which the data and analysis are described.  When authors are not careful in how they describe what they did, this is a major red flag.  Obviously, careful description can mask uncareful data collection, design and analysis, but it is still a signal of credibility that I look to.}{-respondent from USA, political science}



\section{Discussion and Conclusions}

Our study offers an in-depth analysis of researchers' views on reproducibility, replicability, and open science practices in the USA and India, casting light on both commonalities and disparities. While Western nations have more proactively tackled the replication crisis, Indian researchers are becoming more conscious of these issues and are making strides toward embracing open science. 

Yet, our findings indicate that its adoption faces hurdles in both countries and across fields. Respondents across contexts highlight the misalignment of prevailing academic incentives, centered around publications, promotions, and funding, as detrimental to engagement with open science practices. They note a lack of appreciation for replication studies in favor of novelty. We note that new incentives have emerged and should be used as stepping stones for substantial extensions. A notable example is the impact of badges. These simple rewards have real, measurable impact on engagement with open science practices \cite{rowhani2020did, rowhani2018badges}. 
Next steps might include the inclusion of badges into reputational metrics and promotional practices. Journals and conferences might consider dedicated opportunities to publish replication studies. 

Discussions of reproducibility and open science centered in the West have not fully appreciated the challenges faced by researchers working at institutions with fewer resources and less social capital.  For example, when attempting to reproduce or replicate a published finding, respondents in India believe that getting responses from the paper's authors in the West is more challenging for them. 
While it is generally acknowledged that reproduction and replication should ideally be possible without consultation with authors, consultation is still standard practice and ultimately it is a biased practice that favors well-established community members.


Our findings also suggest the potential value of standing up educational and public-facing initiatives in India, e.g., workshops, consortia, centers, where questions about reproducibility and replication have simply received less attention. Yet, our findings also suggest that the mere existence of institutions aiming to improve research practices in India can not be successful without effective execution at the grassroots level. A comprehensive integration of resources, including training and access to necessary resources, is needed to facilitate widespread sharing of research artifacts and open science practices. These resources include but are not limited to: time; computing resources; support for data storage and management; and funding for open-access publication fees. Our study suggests universities in both countries should have undergraduate and graduate courses focused on best practices and highlighting existing challenges.

Globally, establishing metrics to measure engagement with open science and replication efforts could help realign academic incentives. 
Our work confirms shortcoming in researchers' approaches to assessment of confidence in published work. Author reputation was noted as a meaningful signal of credibility in both countries and across domains. Current ways for researchers accumulate good or bad reputation, though, are driven by available existing metrics. The ways in which biases with respect to author reputation may be worsened or mitigated by open science practices, e.g., open peer review, is unclear. 
Finally, our work highlights the urgency of reevaluating existing peer review processes. Ideally, the strongest signal of credibility for a published finding should be the fact that it was peer reviewed and ultimately published. A majority of our respondent expressed dissatisfaction with current peer review processes and suggested that more rigorous mechanisms of research assessment are needed.

Researchers expected to publish rigorous, reproducible, replicable work, adhering to best practices of open science are embedded within social, cultural and economic context. Our work makes clear that this context is not uniform and that solutions which do not consider this will inevitably fall short.  

\bibliographystyle{ACM-Reference-Format}
\bibliography{bibliography}


\end{document}